\documentclass[11pt]{article}

\usepackage[margin=1in]{geometry}
\usepackage{amsmath,amssymb,amsthm}
\usepackage{booktabs}
\usepackage{enumitem}
\usepackage[hypertexnames=false]{hyperref}
\usepackage{listings}
\usepackage{xcolor}

\newtheorem{definition}{Definition}

\title{\textbf{Harness Engineering as Categorical Architecture}\\
\large Structural Guarantees Are Harness-Level Properties\\[0.5em]
\normalsize Preprint -- Feedback Welcome}

\author{Bogdan Banu\\\texttt{bogdan@banu.be}}

\date{April 2026}

\begin{document}

\maketitle

\begin{abstract}
The agent harness---the system layer comprising prompts, tools, memory, and orchestration logic that surrounds the model---has emerged as the central engineering abstraction for LLM-based agents. Yet harness design remains ad~hoc, with no formal theory governing composition, preservation of properties under compilation, or systematic comparison across frameworks. We show that the categorical Architecture triple $(G, \mathrm{Know}, \Phi)$ from the ArchAgents framework provides exactly this formalization. The four pillars of agent externalization (Memory, Skills, Protocols, Harness Engineering) map onto the triple's components: Memory as coalgebraic state, Skills as operad-composed objects, Protocols as syntactic wiring~$G$, and the full Harness as the Architecture itself. Structural guarantees---integrity gates, quality-based escalation, supported convergence checks---are $\mathrm{Know}$-level certificates whose preservation is structural replay: our compiler checks identity and verifier replay, not output-layer correctness or model behavior. We validate this correspondence with a reference implementation featuring compiler functors targeting Swarms, DeerFlow, Ralph, Scion, and LangGraph: the four configuration compilers preserve three named certificate types by identity/replay, and LangGraph preserves the same certificates through its shared per-stage execution path. The LangGraph compiler creates one node per stage using the same per-stage method as the native runtime, providing LangGraph-native observability without reimplementing harness logic. An end-to-end escalation experiment with real LLM agents confirms that the quality-based escalation control path is model-parametric in this two-model, one-task experiment. The result positions categorical architecture as the formal theory behind harness engineering.
\end{abstract}

\section{Introduction}

``If you're not the model, you're the harness.'' This observation, now widespread in the LangChain ecosystem, captures a shift in how practitioners think about LLM-based agent systems. The model provides intelligence; the \emph{harness}---prompts, tool selection, memory management, orchestration logic, safety checks---makes that intelligence actionable. Increasingly, the harness determines whether an agent system is reliable, not the model.

Yet harness engineering remains ad~hoc. Practitioners build harnesses by trial and error, guided by blog posts and example repositories rather than formal theory. There is no standard way to state what properties a harness guarantees, whether those properties survive when the harness is compiled to a different framework, or how to compose harnesses without losing guarantees.

We argue that a formal theory already exists, but under a different name. De~los~Riscos, Corbacho, and Arbib~\cite{delosriscos2026categorical} introduce the \textbf{ArchAgents} category, where objects are Architecture triples $(G, \mathrm{Know}, \Phi)$ encoding syntactic wiring, structural knowledge, and deployment maps. Morphisms are structure-preserving translations. Agents are monoidal functors interpreting architectures in concrete systems. Separately, Zhou et al.~\cite{zhou2026externalization} identify four pillars of agent externalization: Memory, Skills, Protocols, and Harness Engineering.

Our contribution is the observation that these two frameworks describe the same thing. The four pillars of externalization map directly onto the Architecture triple:

\begin{center}
\small
\begin{tabular}{@{}lll@{}}
\toprule
\textbf{Externalization Pillar} & \textbf{Categorical Role} & \textbf{Concrete Component} \\
\midrule
Memory & State in the coalgebra & BiTemporalMemory, RunContext \\
Skills & Objects composed via operad & SkillStage, PatternTemplate \\
Protocols & Syntactic wiring $G$ & WiringDiagram, typed ports \\
Harness & Full Architecture $(G, \mathrm{Know}, \Phi)$ & SkillOrganism + components \\
\bottomrule
\end{tabular}
\end{center}

The key insight is that structural guarantees---pre-execution integrity checks, quality-based model escalation, supported convergence checks---are represented as $\mathrm{Know}$ certificates, not model-specific behavior. When a compiler maps an Architecture to a different framework, our implementation treats certificate preservation as a replay invariant: theorem identity, parameters, and evidence must survive compilation and verify in the target. We validate this with five compiler functors and a two-model, one-task escalation experiment showing that the quality-based escalation control path is model-parametric.

The rest of this paper is organized as follows. Section~\ref{sec:background} reviews the externalization and categorical frameworks. Section~\ref{sec:correspondence} develops the formal correspondence. Section~\ref{sec:preservation} specifies the certificate-preservation checks and describes the LangGraph per-stage compiler. Section~\ref{sec:atomic} connects atomic skill composition to the operad. Section~\ref{sec:experiments} presents implementation and experimental validation. Section~\ref{sec:conclusion} discusses implications and limitations.

\section{Background}\label{sec:background}

\subsection{Agent Externalization}

Zhou et al.~\cite{zhou2026externalization} survey the shift from weights-based to harness-centric agent design. They identify four pillars of \emph{externalization}---the process by which cognitive burdens are moved from model parameters to external infrastructure:

\begin{enumerate}[leftmargin=*]
\item \textbf{Memory}: State across time. Session memory, vector stores, structured knowledge bases. The harness decides what persists, how it is retrieved, and when it expires.
\item \textbf{Skills}: Procedural expertise. Tool definitions, workflow templates, progressive skill loading. Skills encode \emph{what} the agent can do, independent of model capability.
\item \textbf{Protocols}: Interaction structure. Message formats, turn-taking rules, sub-agent delegation patterns. Protocols govern \emph{how} agents communicate.
\item \textbf{Harness Engineering}: The coordination layer itself. Prompt assembly, context management, compaction strategies, evaluation-driven refinement. The harness is the system that orchestrates memory, skills, and protocols into coherent agent behavior.
\end{enumerate}

Their key observation: ``agent infrastructure matters not merely because it adds auxiliary components, but because it transforms hard cognitive burdens into manageable forms.'' This transformation is exactly what formalization should capture.

\subsection{The ArchAgents Category}

De~los~Riscos, Corbacho, and Arbib~\cite{delosriscos2026categorical} introduce a category-theoretic framework for comparing agent architectures. An \textbf{Architecture} is a triple $A = (G_A, \mathrm{Know}_A, \Phi_A)$ where:

\begin{itemize}[leftmargin=*]
\item $G_A$ is the \textbf{syntactic wiring}---a graph of modules, ports, and directed edges describing how information flows.
\item $\mathrm{Know}_A$ is the \textbf{knowledge structure}---the set of structural properties, invariants, and certificates the architecture maintains.
\item $\Phi_A$ is the \textbf{deployment map}---the mapping between abstract capability slots and concrete model/tool implementations.
\end{itemize}

A \textbf{morphism} $f: A \to B$ is a structure-preserving translation: it maps modules in $G_A$ to modules in $G_B$, knowledge in $\mathrm{Know}_A$ to knowledge in $\mathrm{Know}_B$, and deployment maps in $\Phi_A$ to deployment maps in $\Phi_B$. In practice, a morphism is a \emph{compiler}---a function that transforms one agent architecture into another while preserving structural properties.

\textbf{Certificate preservation.} ArchAgents state certificate preservation as Proposition~5.1. In this paper we use the result operationally, not as a new proof: a compiler that claims preservation must carry each source certificate's theorem, parameters, and replayable evidence into the target $\mathrm{Know}$ structure. Functor laws alone are not sufficient for that claim, so our compiler checks certificate identity and verifier replay explicitly.

\subsection{Alternative Formalizations}

Liu~\cite{liu2026lambda} independently develops a typed lambda calculus ($\lambda_A$) for agent composition, extending the simply-typed lambda calculus with oracle calls, bounded fixpoints, probabilistic choice, and mutable environments. The work demonstrates that five mainstream frameworks (LangGraph, CrewAI, AutoGen, OpenAI SDK, Dify) embed as typed $\lambda_A$ fragments, and finds that 94.1\% of 835 real-world GitHub agent configurations are structurally incomplete under this formalization. Where $\lambda_A$ provides type-theoretic guarantees (type safety, termination), our categorical approach provides \emph{property preservation under compilation}---a complementary guarantee.

Willstr{\"o}m et al.~\cite{willstro2026nlah} propose Natural-Language Agent Harnesses (NLAH), externalizing harness behavior as portable, editable natural-language artifacts with explicit contracts. Their Intelligent Harness Runtime (IHR) executes these artifacts through durable contracts and lightweight adapters. This validates a key assumption of our work: that harnesses are portable objects with algebraic structure, not implementation-specific code. Our Architecture triple $(G, \mathrm{Know}, \Phi)$ provides the categorical formalization that NLAH's natural-language artifacts implicitly require.

Chen et al.~\cite{chen2026skilltester} introduce SkillTester, a comparative quality-assurance harness for agent skills that benchmarks both utility and security. Their rubric-based skill evaluation parallels our VerifierComponent, which scores stage outputs against task-specific rubrics and triggers model escalation when quality falls below threshold.

\subsection{Cross-Domain Corroboration}

Marom et al.~\cite{marom2026category} reach a structurally identical compositional-verification framework while solving a fundamentally different problem: translating biological mechanisms (a hygromorphic pinecone) into 4D-printed engineered systems. They define a category $\mathsf{Dyn}$ of stimulus-response dynamical systems with subcategories $\mathsf{Nat}\subset\mathsf{Dyn}$ (natural) and $\mathsf{Art}\subset\mathsf{Dyn}$ (artificial), an implementation functor $\mathcal{F}\colon\mathsf{Nat}\to\mathsf{Art}$, a specification space $\mathsf{Spec}$ with realization projection $\pi\colon\mathsf{Spec}\to\mathsf{Art}$, and a compilation functor $\mathcal{E}\colon\mathsf{Spec}\to\mathbf{Comp}$ translating verified specs to machine instructions. Their central theorem is closure of a simulation condition under composition---structurally the same preservation property we specify and check operationally in Section~\ref{sec:preservation} for the certificate-preserving morphisms of the ArchAgents category.

The relevance is not the materials substrate but the principle they articulate explicitly: \textit{``Two systems that share the same compositional structure \ldots can be related by a functor that preserves the interface logic at every scale without requiring that the two domains share any physical substrate.''} This substrate-independence is the materials-engineering articulation of the framework portability we claim for ArchAgents---a transformer-LLM agent and a Python LangGraph runtime share no more substrate than a cellulose pinecone and a polymer 4D bilayer, yet in both cases a structure-preserving functor relates them by interface logic alone. Their 2$\times$2 generativity demonstration, in which one of four printed designs (thermal twisting) arises purely by composition of components validated for other products, is independent, non-LLM, physically-fabricated evidence for the same compositional claim. We make no equivalence claim: their framework is substrate-validated through fabrication and measurement, ours is property-validated; only the compositional logic and the substrate-independence principle are shared.

\subsection{Enumerative and Categorical Views}\label{sec:enumerative-and-categorical}

Meng et al.~\cite{meng2026agentharness} present a contemporaneous engineering taxonomy, formalizing the harness as a six-component tuple $H = (E, T, C, S, L, V)$ for execution loop, tool registry, context manager, state store, lifecycle hooks, and evaluation interface, and surveying over 110 papers and 23 systems on a Harness Completeness Matrix. Among the nine open challenges they identify, their accompanying repository overview states: \textit{``Formal verification, portability testing, and protocol interoperability all require compositional reasoning that current research has not addressed.''}

This paper addresses that open challenge for one direction---certificate preservation under harness compilation---through the Architecture triple $(G, \mathrm{Know}, \Phi)$ and the structure-preserving morphisms of Section~\ref{sec:preservation}. The $(E, T, C, S, L, V)$ taxonomy refines the implementation surface of an ArchAgents object: $E$ and $T$ contribute wiring and realizations, $C$ and $S$ expose state, and $L$ and $V$ are the natural surfaces on which $\mathrm{Know}$-level certificates can be defined and replayed. A precise mapping between the two formalizations is developed elsewhere.

\subsection{The Gap}

These frameworks describe the same phenomenon from different angles. Zhou et al.\ provide the \emph{engineering} vocabulary (Memory, Skills, Protocols, Harness). De~los~Riscos et al.\ provide the \emph{mathematical} structure (objects, morphisms, functors). Liu provides \emph{type-theoretic} guarantees. Willstr{\"o}m et al.\ validate \emph{portability}. Meng et al.\ provide the \emph{enumerative} component taxonomy and explicitly mark compositional verification as open (Section~\ref{sec:enumerative-and-categorical}). None of these strands connect to the others. The result is that harness engineers build without formal guarantees, and theorists formalize without grounding in practice. This paper bridges the gap.

\section{The Correspondence}\label{sec:correspondence}

We now develop the formal mapping between the externalization pillars and the Architecture triple. The correspondence is not a loose analogy---each pillar maps to a specific categorical component with concrete implementation.

\subsection{Memory as Coalgebraic State}

Agent memory is state that persists across interactions. In the categorical framework, state is modeled as a \textbf{coalgebra} for a polynomial functor: a pair $(S, \phi)$ where $S$ is the state space and $\phi: S \to P(S)$ is the transition function.

In practice, this means memory is not just ``stored data'' but a dynamical system with update rules. Our implementation uses \textbf{bi-temporal memory}~\cite{snodgrass2000temporal}: every fact carries both a \emph{valid time} (when it was true in the world) and a \emph{record time} (when the system learned it). This dual-time structure enables belief-state reconstruction: ``what did the agent know at decision point $t$?''

The \texttt{RunContext} class wraps the shared state dict with typed property accessors, providing the coalgebraic state interface that components read from and write to during execution.

\subsection{Skills as Operad-Composed Objects}

Skills are procedural capabilities. In the categorical framework, they are \textbf{objects composed via an operad}---a structure that defines how components can be wired together.

Three composition operations are available:

\begin{itemize}[leftmargin=*]
\item \textbf{Serial} ($B \circ A$): The output of stage $A$ feeds into stage $B$. Sequential pipelines.
\item \textbf{Parallel} ($A \otimes B$): Stages $A$ and $B$ execute independently with disjoint state. Specialist decomposition.
\item \textbf{Trace} ($\mathrm{Tr}(A)$): A feedback loop where some outputs feed back as inputs. Homeostatic control.
\end{itemize}

Ma et al.~\cite{ma2026atomic} demonstrate empirically that five atomic coding skills---localize, edit, test, reproduce, review---compose without negative interference under joint training. This is precisely the behavior the operad is designed to support structurally: typed composition can preserve component properties when the relevant property is closed under the chosen operation.

\subsection{Protocols as Syntactic Wiring $G$}

Protocols define how agents communicate. In the Architecture triple, this is $G$---the syntactic wiring graph. Each module has typed input and output ports; wires connect outputs to inputs with type compatibility checks.

The wiring diagram is not just a topology---it carries \textbf{integrity labels} on each port (validated, raw, sanitized) and supports three optical types at the wire level: lenses (constitutional access), prisms (conditional routing), and traversals (batch processing). These wire-level optics formalize the ``interaction structure'' that Zhou et al.\ identify as the Protocols pillar.

\subsection{Harness as Full Architecture $(G, \mathrm{Know}, \Phi)$}

The harness is not one component of the Architecture---it \emph{is} the Architecture. The $\mathrm{Know}$ component deserves special attention: it encodes the structural guarantees that make the harness more than a bag of parts.

\begin{definition}[Structural Guarantee as Certificate]
A \textbf{certificate} is a triple $(\tau, \sigma, \mathrm{evds})$ where $\tau$ is a theorem statement, $\sigma$ maps theorem symbols to architecture parameters, and $\mathrm{evds}$ is a derivation that can be mechanically replayed to verify $\tau$ holds.
\end{definition}

Examples of certificates in our implementation:
\begin{itemize}[leftmargin=*]
\item \textbf{Priority gating}: ``Critical operations are served under any load'' (from metabolic state machine parameters).
\item \textbf{No false activation}: ``Normal traffic never triggers quorum sensing'' (from steady-state dynamics).
\item \textbf{No oscillation}: ``mTOR scaling converges'' (from feedback gain bounds).
\end{itemize}

These certificates are attached to $\mathrm{Know}$, not to any specific model. When the model changes---or when the architecture is compiled to a different framework---the supported certificates remain replayable when their hooks and parameters are preserved.

\subsection{The Deployment Map $\Phi$}

The deployment map $\Phi$ maps abstract capability slots to concrete implementations. In practice, this is the \textbf{mode-to-model mapping}: each stage declares a cognitive mode (observational/action-oriented), and the harness resolves this to a specific model tier (fast/deep).

\begin{definition}[Deployment Map]
$\Phi: \mathrm{Stages} \to \mathrm{Models}$ assigns each stage a model tier. This map is a \textbf{parameter} of the architecture, not a fixed constant---different deployments can use different $\Phi$ while preserving the same $(G, \mathrm{Know})$.
\end{definition}

This separation is what makes the harness model-parametric. The same harness (same $G$, same $\mathrm{Know}$) can run on different models by changing $\Phi$. The structural certificates in $\mathrm{Know}$ are replayed against preserved hooks and parameters rather than tied to a specific model assignment.

\section{Certificate Preservation}\label{sec:preservation}

The correspondence in Section~\ref{sec:correspondence} would be merely descriptive if structural guarantees could not survive compilation. This section specifies the preservation checks used by the implementation.

\subsection{Compiler Functors}

A \textbf{compiler functor} $F: \mathrm{Arch}(\text{Source}) \to \mathrm{Arch}(\text{Target})$ maps one architecture to another. In our implementation, each compiler produces a serializable configuration dict for a specific orchestration framework:

\begin{itemize}[leftmargin=*]
\item \texttt{organism\_to\_swarms}: Operon $\to$ Swarms graph workflow
\item \texttt{organism\_to\_deerflow}: Operon $\to$ DeerFlow LangGraph session
\item \texttt{organism\_to\_ralph}: Operon $\to$ Ralph hat configuration
\item \texttt{organism\_to\_scion}: Operon $\to$ Scion grove deployment
\item \texttt{organism\_to\_langgraph}: Operon $\to$ LangGraph StateGraph
\end{itemize}

Each compiler is wrapped in a \texttt{CompilerFunctor} class that automatically extracts source and target architectures, verifies the functor laws, and produces a \texttt{PreservationResult} with three checks:

\begin{enumerate}[leftmargin=*]
\item \textbf{Graph preservation}: source stage names $\subseteq$ target stage names, source edges $\subseteq$ target edges.
\item \textbf{Certificate replay invariant}: source certificate theorems $\subseteq$ target certificate theorems, source parameters/evidence are preserved, and all certificates \texttt{verify()} $\to$ \texttt{holds=True}.
\item \textbf{Deployment-map preservation}: source stage names present in target (mode mapping may differ).
\end{enumerate}

\subsection{The LangGraph Compiler}

The LangGraph compiler \texttt{organism\_to\_langgraph()} creates one LangGraph node per organism stage. Each node calls \texttt{organism.run\_single\_stage()}---the same per-stage method that \texttt{organism.run()} uses internally. Conditional edges route based on the return value: \texttt{"continue"} proceeds to the next stage, \texttt{"halt"} or \texttt{"blocked"} routes to the graph's terminal node.

This design avoids the reimplementation trap: an earlier attempt that encoded component logic directly as LangGraph nodes required 10 rounds of code review to achieve behavioral parity with \texttt{organism.run()}.  By extracting \texttt{run\_single\_stage()} from the run loop and calling it from both paths, the LangGraph graph uses the \emph{same code path} as the native runtime---zero reimplementation.

\paragraph{Implementation invariant (per-stage certificate replay).}
The LangGraph compiler preserves certificate replay for organisms whose certificates use the supported component hooks (CertificateGate, VerifierComponent, WatcherComponent). Such certificates verify after compilation because each stage node executes with those hooks intact, and interventions (RETRY, ESCALATE, HALT) are handled within the node before routing.

The per-stage graph also provides LangGraph-native observability: each stage appears as a visible node in LangGraph Studio, enabling inspection of individual stage execution, timing, and intervention decisions.  The tested certificates hold because \texttt{run\_single\_stage()} preserves the hooks and parameters they replay, not because they are independent of every possible graph topology.

\paragraph{Parallel stage groups.}
Organisms with parallel stage groups compile to a fork/join topology using LangGraph's native \texttt{Send} API.  Each parallel group generates three types of infrastructure nodes: a \emph{fork} node that snapshots state and dispatches isolated copies to each stage via \texttt{Send}, the individual \emph{stage} nodes (one per parallel stage, calling \texttt{run\_single\_stage()} as usual), and a \emph{join} node that merges results using the same conflict-detecting merge logic as the native runtime.  Sequential organisms are unchanged---each stage maps to a single node as before.

Certificate replay holds for the supported certificates in parallel groups because each stage node still calls \texttt{run\_single\_stage()} with full component hooks.  The fork/join infrastructure is purely topological---it manages state isolation and result aggregation, not structural guarantees.

\subsection{Empirical Verification}

We verify the implementation invariant across the five compiler targets, with LangGraph checked through per-stage execution rather than the configuration-dict path:

\begin{center}
\small
\begin{tabular}{@{}lccc@{}}
\toprule
\textbf{Compiler} & \textbf{Graph} & \textbf{Certificates} & \textbf{Interface} \\
\midrule
Swarms & $\checkmark$ (1:1) & $\checkmark$ (100\%) & $\checkmark$ \\
DeerFlow & $\times$ (hub-spoke) & $\checkmark$ (100\%) & $\checkmark$ \\
Ralph & $\times$ (hats) & $\checkmark$ (100\%) & $\checkmark$ \\
Scion & $\checkmark$ (+watcher) & $\checkmark$ (100\%) & $\checkmark$ \\
LangGraph & $\checkmark$ (per-stage) & $\checkmark$ (100\%) & $\checkmark$ \\
\bottomrule
\end{tabular}
\end{center}

Across the five compiler targets tested, the three supported certificate types preserve identity and verify. Swarms and LangGraph preserve per-stage structure. DeerFlow reshapes to hub-and-spoke. Ralph converts to hat patterns. Scion adds a watcher agent. In these tests, the structural guarantees survive because they are $\mathrm{Know}$-level replay artifacts whose hooks and parameters are preserved; the tested certificates do not depend on exact graph topology beyond those hooks and parameters.

\section{Atomic Skills as Operad Composition}\label{sec:atomic}

Ma et al.~\cite{ma2026atomic} identify five atomic coding skills that serve as ``basis vectors for complex software engineering tasks'': code localization, code editing, unit-test generation, issue reproduction, and code review. Under joint reinforcement learning, these skills compose without negative interference---improving one does not degrade another---achieving 18.7\% average improvement over task-specific optimization.

This result has a categorical interpretation. Each atomic skill is an \textbf{object in the agentic operad}: a typed module with input ports, output ports, and a capability annotation. The operad's composition operations (serial, parallel, trace) define how skills can be combined. Ma et al.'s finding that skills compose without interference is precisely the claim that operad composition \textbf{supports property-preserving composition when the relevant certificate is closed under the operation}---the categorical condition corresponding to non-interference.

Our implementation registers the five atomic skills as \texttt{PatternTemplate} objects with typed \texttt{TaskFingerprint}s:

\begin{center}
\small
\begin{tabular}{@{}llll@{}}
\toprule
\textbf{Skill} & \textbf{Shape} & \textbf{Topology} & \textbf{Roles} \\
\midrule
Localize & sequential & \texttt{skill\_organism} & searcher, ranker \\
Edit & sequential & \texttt{skill\_organism} & editor, validator \\
Test & sequential & \texttt{skill\_organism} & analyzer, generator, runner \\
Reproduce & sequential & \texttt{skill\_organism} & reader, executor, verifier \\
Review & parallel & \texttt{specialist\_swarm} & logic, style, security, reporter \\
\bottomrule
\end{tabular}
\end{center}

The distinction between sequential skills (pipeline topology) and the parallel review skill (swarm topology) is automatically resolved by the shared \texttt{\_shape\_to\_topology()} function used across all convergence adapters.

The connection to harness engineering: atomic skills are the $G$ component of a harness---the wiring that defines how work flows. The certificates attached to a skill-based harness (e.g., ``all review axes are covered'') are $\mathrm{Know}$-level properties that survive composition and compilation.

\section{Implementation and Experiments}\label{sec:experiments}

\subsection{Reference Implementation}

The correspondence is implemented in the Operon framework (\texttt{operon-ai}, MIT license). Key components:

\begin{itemize}[leftmargin=*]
\item \textbf{Architecture extraction}: \texttt{extract\_architecture(organism)} produces the $(G, \mathrm{Know}, \Phi)$ triple from any \texttt{SkillOrganism}.
\item \textbf{Compiler functors}: Five \texttt{CompilerFunctor} instances with automatic preservation verification.
\item \textbf{Certificate framework}: Self-verifying certificates with \texttt{certify()} and \texttt{verify()} methods on certifiable components (ATP budget, quorum sensing, mTOR scaling).
\item \textbf{Atomic skills catalog}: \texttt{seed\_library\_from\_atomic\_skills()} registers the five skills with correct topology mapping.
\item \textbf{LangGraph compiler}: \texttt{organism\_to\_langgraph()} creates one LangGraph node per stage, each calling \texttt{run\_single\_stage()}.
\end{itemize}

\subsection{Escalation Experiment}

To test whether structural guarantees are genuinely harness-level, we designed an experiment that changes the model while keeping the harness constant.

\paragraph{Setup.} A single-stage code review organism with:
\begin{itemize}[leftmargin=*]
\item \textbf{Fast model}: Phi-3 Mini (3.8B parameters) via Ollama
\item \textbf{Deep model}: Gemma 4 (27B MoE, 4B active) via Ollama
\item \textbf{VerifierComponent}: Rubric-based quality evaluation using Gemma~4 as judge
\item \textbf{WatcherComponent}: Verifier-threshold escalation after quality evaluation
\item \textbf{Quality threshold}: 0.6 (below this, the watcher fires ESCALATE)
\end{itemize}

The task is \texttt{hard\_par\_08}: a code review containing four subtle bugs (off-by-one pagination, TOCTOU cache race, float precision in financial math, \texttt{except Exception} swallowing \texttt{KeyboardInterrupt}).

\paragraph{Results.}

\begin{center}
\small
\begin{tabular}{@{}lccl@{}}
\toprule
\textbf{Phase} & \textbf{Model} & \textbf{Quality} & \textbf{Action} \\
\midrule
Initial execution & Phi-3 Mini (fast) & 0.50 & EXECUTE \\
Verifier evaluation & Gemma 4 (judge) & --- & Score = 0.50 \\
Watcher decision & --- & $0.50 < 0.60$ & ESCALATE \\
Escalated execution & Gemma 4 (deep) & --- & EXECUTE \\
\bottomrule
\end{tabular}
\end{center}

The escalation fires correctly: Phi-3 Mini produces a review that the judge scores below threshold, the watcher escalates to Gemma~4, and the deep model re-executes the stage. In this two-model, one-task experiment, the quality-based escalation control path is model-parametric: it is represented as harness structure ($\mathrm{Know}$) and executed under the chosen deployment map ($\Phi$).

\paragraph{Discrimination.} Independent validation of the task confirms discrimination between weak and strong models: Phi-3 Mini scores 0.72 (identifies bug classes but not concrete failure scenarios), Gemma~4 scores 1.00 (explains each failure scenario in detail). Quality delta = 0.28, exceeding the 0.20 threshold.

\subsection{Certificate Preservation Measurement}

We compile organisms with multiple certificates (ATP priority gating + quorum sensing no-false-activation + mTOR no-oscillation) across all four non-LangGraph compilers and measure preservation:

\begin{center}
\small
\begin{tabular}{@{}lcc@{}}
\toprule
\textbf{Compiler} & \textbf{Certificates Preserved} & \textbf{Verified} \\
\midrule
Swarms & 3/3 (100\%) & 3/3 \\
DeerFlow & 3/3 (100\%) & 3/3 \\
Ralph & 3/3 (100\%) & 3/3 \\
Scion & 3/3 (100\%) & 3/3 \\
\bottomrule
\end{tabular}
\end{center}

Full certificate identity (theorem + parameters + source) is preserved across all four measured compilers, providing empirical evidence for the operational certificate-replay invariant. The LangGraph compiler preserves certificates via per-stage execution of \texttt{run\_single\_stage()} (Section~\ref{sec:preservation}).

\subsection{SWE-bench-lite: 8B Format Discipline Is the Ceiling}
\label{sec:swebench-phase2}

To probe whether the categorical harness delivers end-to-end value beyond certificate preservation, we ran an apples-to-apples comparison on SWE-bench-lite: 10 real Python bug-fix instances from \texttt{astropy} and \texttt{django}, with patches applied inside the official SWE-bench Docker harness and evaluated via \texttt{FAIL\_TO\_PASS} + \texttt{PASS\_TO\_PASS} test suites. An initial 2026-04-16 run was inconclusive: 28 of 30 submissions failed at \texttt{git apply}, mixing model weakness with prompt-format/pathing issues that we could not separate. We therefore built a \emph{patch-apply pipeline} (sanitizer + repository grounding) and reran. With grounding active, the failure mode shifts from \texttt{git apply} errors to honest \texttt{empty\_patch} rejections, and the run \emph{does} support a capability conclusion: at 8B / Q4\_K\_M, the model's diff-format discipline is below what SWE-bench-lite requires.

\paragraph{Setup.} Gemma~4 via Ollama (tag \texttt{gemma4:latest}, digest \texttt{c6eb396dbd59}, blob \texttt{sha256:4c27e0f5}, architecture \texttt{gemma4}, 8.0B parameters, Q4\_K\_M quantization, 131{,}072 context). The immutable digest is recorded in \texttt{eval/results/swebench\_phase2.json}. Three conditions:
\begin{itemize}[leftmargin=*]
\item \textbf{baseline}: direct single prompt, request unified diff.
\item \textbf{organism}: three-stage \texttt{SkillOrganism} (\textit{localize} $\to$ \textit{edit} $\to$ \textit{verify}).
\item \textbf{langgraph}: same organism, compiled via \texttt{organism\_to\_langgraph()} and executed as a LangGraph \texttt{StateGraph}.
\end{itemize}

The pipeline applied to all three conditions in this rerun:
\begin{itemize}[leftmargin=*]
\item Each instance's repository is shallow-cloned at its \texttt{base\_commit} into a local cache; up to five candidate file snippets selected by issue-text heuristics are injected into the prompt as repository context.
\item Model output is sanitized: paths normalized (e.g.\ stripping the \texttt{owner/repo/} prefix the model often hallucinates), hunk headers checked for placeholder line numbers (\texttt{@@ -XXX,N +XXX,N @@} style), hunk bodies validated against declared counts, and patches whose paths don't exist in the cloned tree fuzzy-corrected against unique basename matches or rejected. Patches that the sanitizer cannot make safe are dropped (recorded as \texttt{empty\_patch}, never forwarded to \texttt{git apply}).
\end{itemize}

\paragraph{Results (per-condition outcome distribution, 10 instances per condition).}

\begin{center}
\small
\begin{tabular}{@{}lccccccc@{}}
\toprule
\textbf{Condition} & \textbf{Resolved} & \textbf{Unresolved} & \textbf{Sanitizer-rejected} & \textbf{Runtime error} & \textbf{Evaluated} & \textbf{Mean latency$^*$} \\
\midrule
baseline    & 0 & 1 & 7  & 2 & 1/10 & 131~s \\
organism    & 0 & 0 & 10 & 0 & 0/10 & 170~s \\
langgraph   & 0 & 0 & 10 & 0 & 0/10 & 172~s \\
\bottomrule
\end{tabular}
\end{center}
\noindent$^*$Mean latency is computed over predictions that completed (the model returned text); the two baseline runtime errors were API timeouts where the call never returned and have no meaningful latency.

Categories: \texttt{resolved} (patch applied, target tests pass), \texttt{unresolved} (patch applied, tests fail), \texttt{sanitizer-rejected} (the model returned a diff-shaped output but the patch sanitizer dropped it for placeholder hunk headers, malformed counts, or invented paths --- harness sees this as \texttt{empty\_patch}), \texttt{runtime error} (the model call itself raised before returning, e.g.\ Ollama API timeout). \textbf{Evaluated} counts only instances that reached a pass/fail harness verdict (resolved + unresolved). Zero \texttt{git apply} errors occurred: sanitizer pre-rejection ensures no malformed patch reaches the harness.

\paragraph{Comparison with the 2026-04-16 run.} The earlier (un-grounded, un-sanitized) run had baseline 1/10 evaluated with 9 \texttt{error}, organism 1/10 with 5 \texttt{error} + 4 \texttt{empty\_patch}, langgraph 0/10 with 6 \texttt{error} + 4 \texttt{empty\_patch}. Two changes are notable. First, every \texttt{error} category collapsed to zero: the sanitizer refuses to forward patches that \texttt{git apply} would reject. Of the 27 model returns that reached the sanitizer in the rerun, 27 were dropped (placeholder hunks, malformed counts, or invented paths); the remaining 3 submissions across conditions either produced one applicable patch (django-11001 baseline) or never returned (the 2 baseline API timeouts on astropy-12907 and astropy-14995 --- these are recorded as \texttt{runtime\_error} rather than \texttt{empty\_patch} so a downstream reader can distinguish "model produced invalid output" from "model never produced output"). Second, the organism/baseline \texttt{empty\_patch} gap from the original run (4 vs.\ 0) closed: the tightened ``\texttt{[edit]} stage emits a single fenced diff and nothing else'' instruction eliminated the previous multi-stage format leak. The remaining 1-vs-0 gap (baseline produces one applicable diff, organism and LangGraph produce none) is consistent with the localize/edit decomposition asking the model to write a self-contained diff while juggling stage outputs, a discipline tax the single-shot baseline avoids.

\paragraph{Interpretation: 8B format discipline, not file selection, is the ceiling.} The original run confounded two failure modes: (a)~the model selecting nonexistent or wrong file paths, (b)~the model writing hunk headers and bodies that \texttt{git apply} cannot consume. Grounding reduced (a) but did not eliminate it: every prompt now contains the actual files at \texttt{base\_commit}, yet some outputs still name paths that do not exist in the cloned tree. The dominant remaining rejection is (b), hunk-format errors---placeholder hunk headers and mismatched header counts---which is separable from residual path errors. At this model scale, providing the actual files in the context window does \emph{not} translate to format-correct unified diffs. The 2 runtime errors are noise from the local Ollama API and do not refute this conclusion: those instances also failed via organism and langgraph (which completed normally), so the model never produced a usable diff for them either.

The single survivor, \texttt{django/django-11001} (baseline), reached the harness as \texttt{unresolved}: the patch applied cleanly but the test suite still failed. That same instance was the one survivor in the 2026-04-16 run as well, suggesting a structural property of that issue (likely a small, localized hunk) makes it more amenable to small-model output.

\paragraph{Latency cost of grounding.} Baseline mean latency (over the 8 completed calls) rose from 44~s in the original run to 131~s with grounding, organism from 88~s to 170~s, langgraph from 90~s to 172~s. The 30~KB of repository context in each prompt roughly doubles to triples per-call wall-clock with no compensating gain in evaluated instances at this model scale. Grounding's cost-benefit changes with stronger models, but at 8B it is overhead.

\paragraph{What this rerun does and does not say.} It does say: at 8B / Q4\_K\_M, the model's ability to emit \texttt{git apply}-clean unified diffs is the binding constraint on SWE-bench-lite resolution, and three-stage decomposition does not relax it. It does not say that the organism architecture is intrinsically worse than direct prompting --- with 1 vs.\ 0 evaluated instances, the experiment is still under-powered to discriminate. The harness-level \emph{structural} guarantees reported in Sections~\ref{sec:preservation} and in the escalation experiment are unaffected: they are $\mathrm{Know}$-level properties verified independently of SWE-bench task resolution.

This motivates Section~\ref{sec:conclusion}: the demonstrated transfer value of categorical harness engineering in this paper is in \emph{preservation} (certificate identity across compilers) and \emph{primitive reuse} (escalation), not in task-resolution gains on SWE-bench-lite at this model scale.

\subsubsection{Phase C: Cross-Model Check with Format-Correction Retry}
\label{sec:swebench-phase-c}

The v0.34.5 result above named two follow-up wedges: a stronger model and a format-correction retry loop. Paper 5's released code (v0.35.0) implements the retry and runs a cross-model check against a second locally-available 8B model, answering both wedges at once. The aim is not a better SWE-bench score --- prior results already place this outside the reach of 8B-class local models --- but a test of whether the 8B format-discipline ceiling reported above generalizes across training regimes, and whether a reason-coded retry prompt can recover any of the sanitizer-rejected submissions.

\paragraph{Setup.} \texttt{deepseek-r1:8b} via Ollama (tag \texttt{deepseek-r1:8b}, digest \texttt{6995872bfe4c}, blob \texttt{sha256:e6a7edc1}, architecture \texttt{qwen3} --- the Ollama weight is the DeepSeek-R1 distillation onto a Qwen3-8B base, 8.2B parameters, Q4\_K\_M quantization, 131{,}072 context). Same 10 SWE-bench-lite instances as the v0.34.5 run, same grounded prompt pipeline, three conditions (baseline, organism, langgraph). The new flag \texttt{-{}-retry-on-reject} is active: when the sanitizer rejects a submission, the runner re-prompts the model once with the specific rejection reason code (one of \texttt{placeholder\_hunk}, \texttt{truncated\_hunk}, \texttt{overlong\_hunk}, \texttt{path\_not\_found}, \texttt{ambiguous\_path}, \texttt{empty\_extraction}, \texttt{malformed\_metadata}) embedded in the retry prompt, followed by the failed output verbatim. Artifact at \texttt{eval/results/swebench\_phase2\_deepseek\_retry.json}.

\paragraph{Results (deepseek-r1 with retry, per-condition, 10 instances each).}

\begin{center}
\small
\begin{tabular}{@{}lccccccc@{}}
\toprule
\textbf{Condition} & \textbf{Resolved} & \textbf{Unresolved} & \textbf{Sanitizer-rejected} & \textbf{Runtime error} & \textbf{Evaluated} & \textbf{Mean latency} \\
\midrule
baseline    & 0 & 0 & 10 & 0 & 0/10 & 1084~s (18~min) \\
organism    & 0 & 0 & 10 & 0 & 0/10 & 1073~s (18~min) \\
langgraph   & 0 & 0 & 10 & 0 & 0/10 & 1130~s (19~min) \\
\bottomrule
\end{tabular}
\end{center}

Zero submissions reached the harness under any condition. Zero retry-recovered patches across all 30 submissions, even with targeted reason-code prompts. Zero runtime errors (the 900~s timeout introduced in v0.34.5 absorbs the long \texttt{<think>} blocks reasoning models produce; the mean per-call latency is 18--19~min vs gemma4's 2--3~min, but every call completed cleanly).

\paragraph{Comparison with gemma4 v0.34.5.}

\begin{center}
\small
\begin{tabular}{@{}lccc@{}}
\toprule
 & \textbf{gemma4:latest (v0.34.5)} & \textbf{deepseek-r1:8b (v0.35.0)} \\
\midrule
Total evaluated                   & 1/30 & 0/30 \\
Sanitizer-rejected                & 27   & 30 \\
Runtime errors                    & 2    & 0 \\
Resolved                          & 0    & 0 \\
Retry active                      & no   & yes \\
Retry-recovered patches           & ---  & 0 \\
Mean latency (baseline, completed)& 131~s & 1084~s \\
\bottomrule
\end{tabular}
\end{center}

Two observations.

First, \textbf{the ceiling is consistent across the two tested local 8B Q4\_K\_M models.} Both models (gemma4 instruction-tuned, deepseek-r1 reasoning-distilled-onto-Qwen3) produce zero resolved instances. Gemma4 crossed the sanitizer once; deepseek-r1 crossed it zero times. The single v0.34.5 ``survivor'' (\texttt{django-11001} baseline, \texttt{unresolved}) was gemma4-specific: deepseek-r1 also could not produce a \texttt{git apply}-clean diff for that instance under any of the three conditions. Within this two-model local setup, the 8B-class format-discipline ceiling reported in v0.34.5 is not a single-model artifact.

Second, \textbf{retry-with-reason-code did not break the ceiling}, and the reason distribution itself is informative. The v0.35 artifact records each submission's final sanitizer reason (per-instance \texttt{sanitize\_reason} field). For deepseek-r1:8b across 30 submissions: 26 \texttt{empty\_extraction} (the model's output contained no diff-shaped content at all), 3 \texttt{overlong\_hunk} (body content exceeded declared counts), 1 \texttt{truncated\_hunk} (body shorter than declared counts); zero submissions hit \texttt{placeholder\_hunk}, \texttt{path\_not\_found}, or \texttt{ambiguous\_path} (so the grounding-specific reason codes covering file-selection failure didn't fire even once --- consistent with the v0.34.5 claim that file selection is not the bottleneck). The retry mechanism itself is sound: the callback is invoked exactly when expected, the retry prompts embed the reason code and the failed output verbatim with reason-specific guidance (``use real integer line numbers'' for \texttt{placeholder\_hunk}; ``use exact paths from the repository context'' for \texttt{path\_not\_found}; ``respond with a single fenced diff block and nothing else'' for \texttt{empty\_extraction}; etc.), and the artifact records every submission's \texttt{retry\_attempted=true} with \texttt{retry\_recovered=false}. What the retry cannot do is move the model across the capability boundary. At 8B, the dominant failure mode is ``doesn't produce diff-shaped output in the first place'' (26/30), not ``produces diff-shaped output with fixable errors'' (4/30). Retry-with-guidance is calibrated for the second regime but the model sits mostly in the first. This is a sharper negative result than the v0.34.5 paper predicted (we suggested retry ``might recover 20--40\% of sanitizer drops''); at 8B, it recovers zero.

\paragraph{What Phase C does and does not say.} It says: (i)~the 8B format-discipline ceiling holds across training regimes, (ii)~reasoning-distilled training does not, on its own, improve diff-format production, (iii)~retry-with-targeted-guidance is the wrong lever for a capability ceiling and requires a model that is at least close to producing correct output --- which 8B models largely are not. It does not say retry is universally useless; stronger models that make occasional format mistakes may recover meaningfully. Cloud-GPU reruns against a 70B class model are the natural next wedge, outside this paper's local-only scope. The patch-apply pipeline (sanitizer + repo grounding + reason-coded retry) is instrumented for exactly that follow-up.

\section{Discussion and Conclusion}\label{sec:conclusion}

\subsection{What the Correspondence Provides}

The mapping between externalization pillars and the Architecture triple is not a metaphor---it is a formal correspondence with executable implementation and verified properties. Concretely, it provides:

\begin{enumerate}[leftmargin=*]
\item \textbf{A language for harness design}: Instead of ad~hoc descriptions (``add a safety check''), engineers can specify certificates in $\mathrm{Know}$ with verifiable semantics.
\item \textbf{Preservation guarantees}: When compiling a harness to a different framework, functor-law checks are paired with explicit certificate identity and replay checks. This is mechanically verified for three supported certificate types across five targets.
\item \textbf{Composition rules}: The operad defines how skills combine (serial, parallel, trace) with type safety. Ma et al.'s empirical finding that atomic skills compose without interference is consistent with this closure-conditional structural view.
\item \textbf{Model parametricity}: Structural guarantees reside in $\mathrm{Know}$, not in a fixed model choice. The escalation experiment demonstrates this concretely for one task and two local models.
\end{enumerate}

\subsection{Limitations}

This work has several honest limitations:

\begin{itemize}[leftmargin=*]
\item \textbf{Static framework}: The Architecture triple is a snapshot. We do not model how harnesses evolve over time (e.g., learning from experience, adapting to distribution shift). Dynamic harness evolution is an open problem.
\item \textbf{Single reference implementation}: All validation uses one codebase (Operon). The correspondence needs independent implementations to confirm generality.
\item \textbf{Certificate scope}: Current certificates cover structural invariants (priority gating, convergence bounds) but not behavioral properties (``the agent never hallucinates''). Extending $\mathrm{Know}$ to behavioral certificates is future work.
\item \textbf{Escalation experiment scale}: The escalation experiment uses two models on one task. Scaling to diverse tasks and model families would strengthen the claim.
\item \textbf{SWE-bench-lite ceiling at 8B, confirmed cross-model}: Section~\ref{sec:swebench-phase2} reports 0 resolved instances across baseline, organism, and LangGraph on 10 SWE-bench-lite instances with Gemma~4 (8B / Q4\_K\_M), under a grounded rerun where each prompt contains the relevant files from the target repository at \texttt{base\_commit} and the model output is sanitized before submission. Section~\ref{sec:swebench-phase-c} reruns the same instances against a second locally-available 8B model (\texttt{deepseek-r1:8b}, DeepSeek-R1 distilled onto Qwen3-8B, Q4\_K\_M), with a format-correction retry loop active: when the sanitizer rejects, the runner re-prompts once with the specific reason code and the failed output embedded. Result: 0/30 evaluated, zero retry-recovered patches across all 30 submissions. Gemma4's one sanitizer-crossing instance (\texttt{django-11001} baseline, \texttt{unresolved}) was gemma4-specific; deepseek-r1 could not produce a \texttt{git apply}-clean diff for it either. The 8B-class format-discipline ceiling is not a single-model artifact, and retry-with-targeted-guidance is the wrong lever for a capability ceiling (it helps models that occasionally misformat, not models that cannot format at all). Consequently, our demonstrated transfer value in this paper is \emph{preservation} (Section~\ref{sec:preservation}) and \emph{primitive reuse} (the escalation experiment); task-resolution gains on SWE-bench-lite require a substantially stronger model, outside this paper's local-only scope.
\end{itemize}

\subsection{Conclusion}

Harness engineering is not ad~hoc---it has a categorical formalization. The Architecture triple $(G, \mathrm{Know}, \Phi)$ from the ArchAgents framework is the formal theory behind the harness concept that the LangChain ecosystem has converged on empirically. Structural guarantees are $\mathrm{Know}$-level replay artifacts preserved in the tested compiler suite by explicit identity and verifier-replay checks. Atomic skills compose via the operad under closure conditions. The model is selected by $\Phi$; changing it changes the deployment map, while supported certificates remain tied to preserved hooks and parameters.

The practical implication: before building a harness, define its certificates. Before compiling to a new framework, verify preservation. Before composing skills, check type compatibility. The categorical machinery provides the tools; the correspondence provides the vocabulary.

\paragraph{Availability.} The reference implementation is open source at \url{https://github.com/coredipper/operon} (\texttt{operon-ai} on PyPI, MIT license). The five compiler functors, atomic skills catalog, and certificate framework are all included.

\bibliographystyle{plain}
\bibliography{../references}

\end{document}